\documentclass[prd,reprint,superscriptaddress,nofootinbib,longbibliography]{revtex4-1}

\usepackage[utf8]{inputenc}
\usepackage[T1]{fontenc}
\usepackage[english]{babel}
\usepackage{amsmath,amssymb,bm,physics}
\usepackage{graphicx}
\usepackage{epstopdf}
\usepackage{color,xcolor}
\usepackage{slashed}
\usepackage{hyperref}
\hypersetup{colorlinks=true,linktoc=page,linkcolor=blue,citecolor=red,urlcolor=cyan}

\newcommand{\hu}{\hspace{0.6cm}}

\newcommand{\dime}{d}

\usepackage{xparse}
\NewDocumentCommand{\dkpara}{O{k} O{\dime}}{\frac{{\rm d} {#1}^{\parallel}}{(2\pi)^{#2}}}
\NewDocumentCommand{\dkd}{O{k} O{\dime}}{ \frac{{\rm d}^{#2} {#1}}{(2\pi)^{#2}}}
\NewDocumentCommand{\dxd}{O{x} O{\dime}}{ {\rm d}^{#2} {#1} }
\NewDocumentCommand{\ad}{O{} O{}}{ \bigl \langle {#1}\bigr\rangle_{{\rm ad}{#2}} }

\begin{document}

\noindent\makebox[\textwidth][r]{\small TUM-HEP-1577/25}
\vspace*{\baselineskip}

\title{Heat Kernels and Resummations: the Spinor Case}

\author{S.~A.~Franchino-Viñas}
\affiliation{Departamento de F\'isica, Facultad de Ciencias Exactas Universidad Nacional de La
Plata, C.C. 67 (1900), La Plata, Argentina}
\affiliation{CONICET, Godoy Cruz 2290, 1425 Buenos Aires, Argentina}
\affiliation{Universit\'e de Tours, Universit\'e d'Orl\'eans, CNRS, Institut Denis Poisson, UMR 7013, Tours, 37200, France
}

\author{C.~García-Pérez}
\affiliation{DIME, Universit\`a di Genova, Via all'Opera Pia 15, 16145 Genova, Italy}
\affiliation{INFN Sezione di Genova, Via Dodecaneso 33, 16146 Genova, Italy}

\author{F.~D.~Mazzitelli}
\affiliation{Centro At\'omico Bariloche,  CONICET,
Comisi\'on Nacional de Energ\'\i a At\'omica, R8402AGP Bariloche, Argentina}
\affiliation{Instituto Balseiro, Universidad Nacional de Cuyo, R8402AGP Bariloche, Argentina}

\author{S.~Pla}
\affiliation{Physik-Department, Technische Universit\"at M\"unchen, James-Franck-Str., 85748 Garching, Germany}

\author{V.~Vitagliano}
\affiliation{DIME, Universit\`a di Genova, Via all'Opera Pia 15, 16145 Genova, Italy}
\affiliation{INFN Sezione di Genova, Via Dodecaneso 33, 16146 Genova, Italy}

\begin{abstract}
Among the available perturbative approaches in quantum field theory, heat kernel techniques provide a powerful and geometrically transparent framework for computing effective actions in nontrivial backgrounds. In this work, resummation patterns within the heat kernel expansion are examined as a means of systematically extracting nonperturbative information. Building upon previous results for Yukawa interactions and scalar quantum electrodynamics, we extend the analysis to spinor fields, demonstrating that a recently conjectured resummation structure continues to hold. The resulting formulation yields a compact expression that resums invariants constructed from the electromagnetic tensor and its spinorial couplings, while preserving agreement with known proper-time coefficients. Beyond its immediate computational utility, the framework offers a unified perspective on the emergence of nonperturbative effects (such as Schwinger pair creation) in relation to perturbative heat kernel data, and provides a basis for future extensions to curved spacetimes and non-Abelian gauge theories.
\end{abstract}

\maketitle

\section{Introduction}\label{sec:intro}

Tunnelling phenomena occupy a central role in quantum mechanics and quantum field theory, providing direct access to processes that lie beyond conventional perturbation theory. From the decay of metastable states to barrier penetration, semiclassical configurations reveal exponential contributions to observables that perturbative expansions alone cannot capture. Such instanton-mediated effects span a wide range of physical settings: in strong-field quantum electrodynamics, the Schwinger effect manifests as electron–positron pair production via tunnelling in an external electric field \cite{Affleck:1981bma,Dunne:2005sx}, while in gravitational physics, Hawking radiation may be interpreted as particles tunnelling across a black hole horizon \cite{Parikh:1999mf}. More broadly, instantons govern processes such as vacuum decay, anomaly generation, and nonperturbative corrections to effective actions \cite{Coleman1985,Callan:1977pt,Belavin:1975fg}, illustrating how semiclassical trajectories in Euclidean spacetime encode exponentially suppressed contributions. Understanding the origin and structure of these effects remains essential for a complete description of quantum dynamics and motivates the search for further methods capable of extracting nonperturbative information, even those that at first sight appear to be purely perturbative.

Heat kernel techniques, a cornerstone of spectral geometry and quantum field theory, offer one such route. The heat kernel short-time asymptotic expansion encodes local geometric information through a hierarchy of coefficients that describe the properties of the underlying manifold and field content \cite{Vassilevich:2003xt,Gilkey1995}. While this expansion is conventionally regarded as a perturbative tool, certain sequences of terms can be reorganised or \textit{resummed}, giving rise to structures that reveal hidden nonperturbative behaviours~\cite{Dunne:2012ae,Voros:1986vw,Avramidi:2009quh}. This insight establishes a conceptual bridge between the semiclassical intuition of tunnelling processes and the analytic machinery of the heat kernel formalism, allowing nonperturbative features to emerge clearly from perturbative data.

At the methodological level, resummation has been applied in multiple contexts. In first-quantised settings, the heat kernel can be interpreted as a propagator in imaginary time, allowing for a resummation over the potential \cite{Wigner:1932eb}. In quantum field theory, covariant perturbation theory provides systematic tools for backgrounds with rapidly varying curvatures or potentials, enabling, for instance, the computation of beta functions in a momentum-like scheme \cite{Barvinsky:1987uw, Barvinsky:1990up, Gorbar:2002pw, Gorbar:2003yt, Franchino-Vinas:2018gzr,Silva:2023lts}. These general strategies naturally lead to more specific applications in curved spacetime: the resummation of the Ricci scalar \cite{Parker:1984dj,Jack:1985mw} has played a key role in analysing how fermionic and scalar condensates respond to background geometry \cite{Flachi:2014jra, Flachi:2015sva, Flachi:2019btk, Castro:2018iqt, Flachi:2025gxo}, while similar techniques have clarified Casimir self-interactions under spacetime-dependent boundary conditions \cite{Franchino-Vinas:2020okl, Edwards:2021cyp, Ahmadiniaz:2022bwy}. For sufficiently simple backgrounds, closed expressions can even be obtained for the effective action \cite{Heisenberg:1936nmg, Weisskopf:1936hya, Brown:1975bc} or the heat kernel itself \cite{Avramidi:1995ik, Avramidi:2009quh}.

Building on these ideas, we have recently initiated a program aimed at establishing the existence of further resummation patterns for the diagonal elements of the heat kernel and for the effective action of quantum fields interacting with nontrivial backgrounds~\cite{Navarro-Salas:2020oew,Franchino-Vinas:2023wea,Franchino-Vinas:2024jvc}. In our previous works, we focused on certain strong-field resummations for scalar Yukawa interactions and for a quantum scalar field coupled to a vector background. In the present paper, we extend this analysis by exploring possible generalisations of the heat kernel resummation ansatz to systems involving spinor fields. Our goal is to outline a unified framework in which perturbative expansions, when appropriately reorganised, already encode the essential elements of some pieces of nonperturbative physics.v

Unless otherwise stated, all computations are performed on a $d$-dimensional Euclidean background metric, with the understanding that a Wick rotation connects the results to their Minkowskian counterparts.

\section{Effective action and heat kernel methods}

The generating functional for scattering amplitudes of a quantum field theory system is defined as
\begin{equation}\label{eq:generating_functional}
    Z[J] := \int\mathcal{D}\varphi~\exp(-S[\varphi] + J\varphi)\,,
\end{equation}
where $S[\varphi]$ denotes the classical action functional, and the path integral runs over all admissible field configurations. Here, $\varphi$ is treated as a generic field, without specifying any internal or spacetime indices. From the generating functional of connected Green’s functions, $W[J]$, defined by
\begin{equation}
    e^{W[J]}:= Z[J]\,,
\end{equation}
the effective action $\Gamma$ is  constructed through a Legendre transform,
\begin{equation}
    \Gamma[\phi] := W[J] - \int d^\dime x~J\phi\,,
\end{equation}
where $\phi$ represents the expectation value of the quantum operator associated with $\varphi$. By definition, $\Gamma$ satisfies thus an implicit equation in terms of the functional integral in Eq.~\eqref{eq:generating_functional}, which is generally not solvable analytically. However, following the techniques presented by Schwinger and DeWitt \cite{DeWitt:1965}, it is possible to show that, for systems whose action is quadratic,
\begin{equation}
    S[\varphi]:= \int {\rm d}^\dime x~\bar{\varphi}\,\mathcal{Q}\,\varphi\,,
\end{equation}
with $\mathcal{Q}$ a differential operator, the functional integral becomes Gaussian and admits an explicit solution, 
\begin{align}\label{eq:log_det}
    \Gamma[\phi] &= S[\phi] +  c_s\Gamma_1:=S[\phi] +  c_s\log\operatorname{Det}\mathcal{Q}\,,
\end{align}
Here, the first term corresponds to the classical action, while the second term defines the quantum part of the effective action, $\Gamma_1$, whose coefficient $c_s$ is a real-valued constant depending on the spin of the quantum field.
$\Gamma_1$ is the primary object of study in this work and it is important to emphasise that, whenever the action incorporates interaction terms, i.e. it goes beyond the quadratic case, the result in Eq.~\eqref{eq:log_det} becomes a one-loop approximation, which is still a highly nontrivial quantity.

A convenient representation of the one-loop effective action is provided by the heat kernel operator~\cite{Vassilevich:2003xt},
\begin{equation}\label{Kdef}
    K(\tau):=\exp(-\tau\mathcal{Q})\,,
\end{equation}
allowing one to write
\begin{align}
    \Gamma_1 
    &= -\int_0^{\infty} \frac{ {\rm d} \tau}{\tau}\Tr K(x,x';\tau) 
    \\
    &= -\int_0^{\infty} \frac{ {\rm d}\tau}{\tau} \int {\rm d}^{\dime}x \ K(x,x;\tau)\,,
\end{align}
with $x'^\mu$ being an arbitrary reference point in spacetime. From the definition of the heat kernel, it can be shown that the elements $K(x,x';\tau) := \langle x|K(\tau)|x'\rangle$ in this expression satisfy the following differential equation and initial condition:
\begin{equation}\label{HKeq}
    (\partial_\tau + \mathcal{Q})K(x,x';\tau) = 0,\qquad K(x,x';0^+) = \delta(x-x')\,.
\end{equation}

In general, Eq.~\eqref{HKeq} is solved perturbatively using a proper-time expansion of the heat kernel matrix elements,
\begin{align}
    K(x,x';\tau)=\sum_{j=0}^{\infty} b_j(x,x')\, \tau^{j-d/2}\,.
\end{align}
This method provides a recursive procedure to compute the Gilkey--Seeley--DeWitt (GSDW) coefficients ($b_j$); it thus offers a perturbative approach to calculating both the heat kernel and, by extension, the effective action. While widely applicable, this approach does not come without some shortcomings: for starters, the recursive calculation of the GSDW coefficients becomes increasingly cumbersome to calculate for every new step, quickly turning into a computationally taxing problem for any except the most simple systems. More importantly, the perturbative nature of the expansion can obscure nonperturbative features of the system.  

To circumvent these issues, as partially discussed in Sec.~\ref{sec:intro}, several works have developed resummed formulations of the heat kernel. In the context of first quantisation, the heat kernel acts as a propagator in imaginary time, for which a resummation of the potential has been established~\cite{Guven:1986gi}. In curved spacetime, another contribution that has been resummed is the Ricci scalar  for the case of a system consisting of a quantum field minimally coupled to a classical gravitational background~\cite{Jack:1985mw,Parker:1984dj,Calzetta:1986pj}. More recently, resummed expansions for the fundamental invariants $\mathcal{F} := F_{\mu\nu}F^{\mu\nu}$ and $\mathcal{G} := \tilde{F}_{\mu\nu} F^{\mu\nu}$ in (S)QED have been conjectured and partially proven~\cite{Navarro-Salas:2020oew,Franchino-Vinas:2023wea}. The present work continues and extends these results, applying resummation techniques to a wider class of systems and demonstrating how nonperturbative information can be extracted directly from perturbative expansions.

\section{Resummation techniques - the case of a spinor in an electromagnetic background}
\label{sec:3}

The starting point for our analysis comes from yet another classic resummation scheme. It has been previously shown that, for any system consisting of a quantum scalar field interacting with an at most quadratic potential background, i.e. for a system where it is possible to write
\begin{equation}\label{Q-BD}
    \mathcal{Q}_{\rm quad} = -\partial^2 + \alpha + \beta_\mu(x-x')^\mu + \frac{1}{4}\gamma^2_{\mu\nu}(x-x')^\mu(x-x')^\nu\,,
\end{equation}
where $\alpha$, $\beta$, and $\gamma$ are constant coefficients and $x'$ is chosen to be the same as in Eq.~\eqref{HKeq}, a fully explicit, closed expression for the heat kernel and the one-loop effective action can be derived~\cite{Brown:1975bc}. This expression has been recently extended in Ref.~\cite{Franchino-Vinas:2023wea} to include systems with arbitrary potential backgrounds, effectively resumming all the contributions made only of the potential and its first two derivatives out of the proper time expansion,  thereby greatly simplifying the expressions of the corresponding generalised GSDW coefficients [see Eq.~\eqref{eq:Omega} below]. This general framework was then applied to a scalar QED system, where it was shown to allow the resummation of the $\mathcal{F}$ and $\mathcal{G}$ scalars defined above.

Now consider the action for a single quantum Dirac spinor $\Psi$ interacting with a classical electromagnetic background,
\begin{align}
    S := -\int {\rm d}^\dime x~\bar{\Psi}(\gamma^{\mu} D_\mu + m)\Psi, \qquad D_\mu := \partial_{\mu} + i e A_{\mu}\,,
\end{align}
where $A_\mu$ denotes the electromagnetic vector potential, $e$ is the coupling constant (typically identified with the field’s charge), $m$ is the mass of the spinor and $\gamma^\mu$ are the $d$-dimensional Dirac matrices (we will always assume that $d$ is even, so that a unique representation of the $\gamma$ matrices exists, but otherwise leave $d$ arbitrary). The operator for which we wish to find a heat kernel resummation formula in this case is then given by\footnote{In the following we will omit the identity matrix in the spinor bundle.}
\begin{align}\label{Q}
    &\mathcal{Q}_{\rm EM}  := \left(\gamma^\mu D_\mu + m\right) \left(-\gamma^\nu D_\nu + m\right)\nonumber \\ 
    &  = 
    -\partial^2 - ie \left(2A^\mu\partial_\mu + \partial_\mu A^\mu\right) + m^2 + e^2A^2  - \frac{i}{2}e\sigma^{\mu\nu}F_{\mu\nu}\,.
\end{align}

Let us first consider the case of a constant and homogeneous electromagnetic field, $F_{\mu\nu}(x)\to \bar F_{\mu\nu}$. In this case, we can find a straightforward expression for the vector potential
\begin{equation}\label{AFc}
    A_\mu(x) = -\dfrac{1}{2} \bar F_{\mu\nu}(x-x')^\nu\,.
\end{equation}
For ease of notation, we will denote $\bar{x}^\mu:=(x-x')^\mu$ from now on. By introducing (\ref{AFc}) into (\ref{Q}), we arrive at\footnote{We treat $F^{\mu}{}_{\nu}$ as a matrix in its spacetime indices.}
\begin{align}\label{QFc}
    \mathcal{Q}_{\rm h} := -\partial^2 &+ ie \bar F^{\mu\nu}\bar{x}_\mu\partial_\nu + m^2 \nonumber\\
    &- \frac{i}{2}e  \sigma^{\mu\nu}\bar F_{\mu\nu} + \frac{1}{4}e^2(\bar F^2)_{\mu\nu}~\bar{x}^\mu\bar{x}^\nu\,,
\end{align}
and we can make direct contact with the general expression in (\ref{Q-BD}) by setting
\begin{align}
\begin{split}
        \alpha_{\rm h} & := m^2 - \frac{i}{2}e\sigma^{\mu\nu}\bar F_{\mu\nu}\,,\\
    \beta^\mu_{\rm h} & := 0\,,\\
    (\gamma_{\rm h}^2)_{\mu\nu} & := e^2(\bar F^2)_{\mu\nu}\,.
\end{split}
\end{align}

The term linear in derivatives in \eqref{Q}, absent from \eqref{Q-BD}, does not modify the computation of the heat kernel. To demonstrate this, we shall initially disregard this term and proceed following the method of Brown and Duff \cite{Brown:1975bc} to obtain a closed-form expression for the heat kernel,
\begin{align}\label{KFc}
K_{\rm h}(x,x';\tau) := \frac{1}{(4\pi\tau)^{\dime/2}}\frac{e^{-\tau\alpha_{\rm h}-\frac{1}{4}\bar{x}^\mu \mathcal{A}_{\mu\nu}^{-1}(\tau)\bar{x}^\nu -\mathcal{C}(\tau)}}{ \det^{1/2}\big(\tau^{-1} \mathcal{A}(\tau) \big)},
\end{align}
where we have defined the functions
\begin{align}
\begin{split}
    \mathcal{A}_{\mu\nu}(\tau):&=\left[ \frac{1}{\gamma_{\rm h}}\tanh(\gamma_{\rm h} \tau)\right]_{\mu\nu}\,, \\
    \mathcal{C}(\tau):&=\tfrac{1}{2}\left[\log\bigl(\cosh(\gamma_{\rm h} \tau)\bigr)\right]^{\mu}{}_{\mu}\,.
\end{split}
\end{align}

By expanding the exponential, it is evident that the heat kernel in (\ref{KFc}) contains only even powers of $\bar{x}$. When we act on this expansion with the linear-derivative term in (\ref{QFc}), we generate contributions of the form $\bar{x}_\mu(F^{2k+1})^{\mu\nu}\bar{x}_\nu$, which vanish identically due to the antisymmetry of $F_{\mu\nu}$. Consequently, our solution in Eq.~\eqref{KFc} satisfies the equation~\eqref{HKeq} both with and without the linear-derivative term. Since the heat kernel equation admits a unique solution, this term carries no additional information and can therefore be consistently neglected. In the coincidence limit, this result precisely reproduces the well-known Euler--Heisenberg expression.

For a general electromagnetic background, we would like to generalise Eq.~\eqref{AFc} and write $A^\mu$ in terms of gauge-invariant quantities. We can do so by choosing the Fock--Schwinger gauge, defined by
\begin{equation}
    \bar{x}^\mu A_\mu(x) = 0\,,
\end{equation}
which allows us to write the potential $A_\mu$ in terms of $F_{\mu\nu}$ and its derivatives at the point $x'$, see Ref.~\cite{pascual1984qcd}:
\begin{equation}
    A_\mu(x) = \sum_{k=0}^{\infty}\frac{1}{k!(k+2)}\bar{x}^{\mu_1}...\bar{x}^{\mu_k}\bar{x}^{\rho}~\partial_{\mu_1...\mu_k}F_{\rho\mu}(x')\,.
\end{equation}
Plugging back this relation for the electromagnetic potential into the operator, we find an expression that resembles~\eqref{Q-BD}, summed to higher powers of $\bar x$ (or higher derivatives of $F_{\mu\nu}$). This time, however, the coefficients take a more complicated form,
\begin{align}\label{abg}
\begin{split}
    \alpha & := m^2 - \frac{i}{2}e\sigma^{\rho\lambda}F_{\rho\lambda}(x')\,, \\
    \beta_\mu & := -ie \left(\frac{1}{3}\partial^\rho F_{\mu\rho}(x') + \frac{1}{2}\sigma^{\rho\lambda}\partial_\mu F_{\rho\lambda}(x')\right)\,, \\
    (\gamma^2)_{\mu\nu} & := e^2 \big(F^2\big)_{\mu\nu}(x') +  \\
&\hspace{1cm}+ie\left(\partial^\rho\partial_{(\mu} F_{\nu)\rho}(x') + \sigma^{\rho\lambda}\partial_\mu\partial_\nu F_{\rho\lambda}(x')\right)\,,
\end{split}
\end{align}
where we have denoted idempotent symmetrisation of indices by enclosing them in parenthesis. 
We now state explicitly the resummation formula that we will prove. Our claim is that the heat kernel takes the form
\begin{small}
\begin{align}\label{HKF}
&K_{\rm EM}(x,x';\tau) :=\nonumber\\
&\frac{1}{(4\pi\tau)^{\dime/2}}\frac{e^{-\tau\alpha-\frac{1}{4}\tilde\sigma^\mu(x,x') \mathcal{A}^{-1}_{\mu\nu}(x';\tau)\tilde\sigma^{\nu}(x,x') -\mathcal{C}(x';\tau)}}{ \det^{1/2}\left(\tau^{-1} \mathcal{A}(x'; \tau) \right)}\Omega(x,x';\tau)\,,
\end{align}
\end{small}
where the auxiliary functions are defined as
\begin{align}\label{HKFcoef}
\begin{split}
    \tilde\sigma_\mu(x,x'):&= \bar{x}_\mu + \mathcal{B}_\mu(x';\tau), \\ \mathcal{A}_{\mu\nu}(x;\tau):&=\left[ \frac{1}{\gamma}\tanh(\gamma \tau)\right]_{\mu\nu}, \\
    \mathcal{B}_{\mu}(x;\tau):&= 2 \beta^\nu \left[\gamma^{-2}\Big( 1-\sech(\gamma \tau)\Big)\right]_{\nu\mu}, \\
    \mathcal{C}(x,\tau):&=  \beta^\mu \left[-\tau \gamma^{-2} +\gamma^{-3}\tanh(\gamma \tau)\right]_{\mu\nu} \beta^\nu \\
    &\hu+\tfrac{1}{2}\left[\log\bigl(\cosh(\gamma \tau)\bigr)\right]^{\mu}{}_{\mu}\,,
\end{split}
\end{align}
and $\alpha$, $\beta$, and $\gamma$ are given by the expressions in Eq.~\eqref{abg}. The function $\Omega(x,x';\tau)$ admits a proper-time expansion\footnote{In flat space, the coefficient $a_0(x,x')$ coincides with the parallel transport operator $\mathcal{P}(x,x')$, the path-ordered exponential of the gauge connection along the geodesic joining $x$ to $x'$ \cite{Barvinskii:2024iqz}. In our case, the parallel transport becomes trivial along the gauge direction, yielding $\mathcal{P}(x,x')=\mathbf{1}$.}
\begin{equation}\label{eq:Omega}
    \Omega(x,x';\tau) = \sum_{j=0}^{\infty}a_j(x,x')\, \tau^j\,, \qquad a_0(x,x') = 1\,.
\end{equation}
The remarkable property of this resummation is that, as we are going to prove below, when the coincidence limit $x'\rightarrow x$ of the coefficients is taken, none of them depend on any of the electromagnetic invariants contained in the set
\begin{equation}\label{chains}
    \mathcal{K}:=\{\left(\sigma^{\rho\lambda}F_{\rho\lambda}\right)^j, (F^j)^\mu_{~~\mu},\, j> 0\}\,.
\end{equation}

This result shows that our resummation effectively removes all dependence on the electromagnetic field strength invariants from the coincidence limit of the heat kernel coefficients, i.e. we have at disposal a non-perturbative resummation of the electromagnetic background effects. This property is crucial for applications to Schwinger pair production and related strong-field phenomena.

We shall sketch the proof of our claim by explicitly deriving the recurrence relation satisfied by the coefficients $a_j(x,x')$, which arises from substituting expression~\eqref{HKF} into the heat kernel equation \eqref{HKeq}, and grouping all resulting terms in powers of the proper time $\tau$. The result, for every $j\geq 0$, is
\begin{widetext}
\begin{align}\label{HKrec-eq}
\begin{split}
-\left(j+1+\bar{x}_\alpha \partial^\alpha \right) a_{j+1}(x,x') = &~ (-\partial^2 +\mathfrak{S}) ~a_{j}(x,x') + 2A^\mu\partial_\mu a_j(x,x') \\
&\hspace{0.25cm}+ \sum_{n=1}^{\lfloor \frac{j}{2} \rfloor} \frac{B_{2n}}{(2n)!} \Big( 4(2^{2n}-1) \beta^\alpha \left(\gamma^{2(n-1)}\right)_{\alpha\beta} + 2^{2n}\bar{x}^\alpha\left(\gamma^{2n}\right)_{\alpha\beta}\Big) \partial^\beta a_{j+1-2n}(x,x') \\
&\hspace{0.5cm}- \sum_{n=1}^{\lfloor \frac{j+1}{2} \rfloor} \frac{B_{2n}}{(2n)!} \Big( 4(2^{2n}-1) \beta^\alpha \left(\gamma^{2(n-1)}\right)_{\alpha\beta} + 2^{2n}\bar{x}^\alpha\left(\gamma^{2n}\right)_{\alpha\beta}\Big) A^\beta a_{j+1-2n}(x,x')\,,
\end{split}
\end{align}
\end{widetext}
where $B_{k}$ denotes the $k$th Bernoulli number, $\lfloor\cdot\rfloor$ is the floor function, and we define the effective potential $\mathfrak{S}$ as
\begin{align}
\mathfrak{S}:&= m^2-\frac{i}{2}\sigma^{\alpha\beta}F_{\alpha\beta}-ie\partial_\mu A^\mu + e^2A^2 \nonumber\\ 
&\hspace{0.5cm}
-  \alpha -  \bar{x}^\alpha \beta_\alpha -\frac{1}{4}  \bar{x}^\alpha\bar{x}^\beta (\gamma^2)_{\alpha\beta}\,.
\end{align}

There are three key ingredients to this proof. The first is that, given its particular form, $\mathfrak{S}$ and its first two derivatives vanish identically in the coincidence limit, i.e.
\begin{equation}
    \lim_{x'\rightarrow x} \mathfrak{S} = \lim_{x'\rightarrow x}\partial_\mu\mathfrak{S} = \lim_{x'\rightarrow x}\partial_\mu\partial_\nu\mathfrak{S} = 0\,.
\end{equation}
By carefully analyzing the structure of $\mathfrak{S}$, we see that this property ensures that all contributions from the effective potential in the coincidence limit depend solely on derivatives of $F_{\mu\nu}$. 

The second ingredient is the particular dependence of the coefficients $\alpha,\beta,\gamma$, cf. \eqref{abg}, as well as the potential $A^\mu$, on the electromagnetic field. While $\alpha$ has already been accounted for in the previous discussion (appearing only inside $\mathfrak{S}$), and $\beta$ depends only on derivatives of $F_{\mu\nu}$, the coefficient $\gamma$ has the structure
\begin{equation}
    \gamma^2 \sim F^2 + \text{terms with derivatives of }F,
\end{equation}
meaning that its only contributions without derivatives of $F_{\mu\nu}$ are of the form $(F^{2j})_{\mu\nu}$. As for $A^\mu$, most of its contributions involve derivatives of $F_{\mu\nu}$, the exception being the leading order, which is identical to the constant electromagnetic field case, cf. Eq.~\eqref{AFc}. 

Returning to Eq.~(\ref{HKrec-eq}), we see that all of this implies no element of the set $\mathcal{K}$ explicitly appears in the recurrence relation. Their only possible appearances could come from the coefficients $a_k(x,x')$ themselves, or from contractions arising when taking derivatives of terms involving $\bar{x}$. However, the third ingredient, the fact that $a_0(x,x')=1$, allows us to construct a proof by induction showing this does not occur either. 

Indeed, an explicit ordering is induced by the recurrence relation, $a_0\rightarrow a_1\rightarrow\partial_\mu a_1\rightarrow\partial_\mu\partial_\nu a_1\rightarrow a_2...$, where to calculate a given element all the previous are needed. Taking the appropriate derivatives of Eq.~\eqref{HKrec-eq}, one can verify that the diagonal of a given term in this sequence would contain an object in $\mathcal{K}$ only if a previous one does. Since the first element manifestly does not, we conclude that none of the diagonal coefficients $a_k(x,x)$, nor any of the diagonal derivatives $\partial_{\mu_1}\cdots\partial_{\mu_n} a_k(x,x)$, can depend on the invariants built exclusively from $F_{\mu\nu}$ and $\sigma_{\mu\nu}$.

The full dependence of the heat kernel on these objects is therefore completely contained within the global prefactor in (\ref{HKF}), and is valid for an arbitrary even dimension $d$. In particular, this completes and expands the proof, initiated in \cite{Franchino-Vinas:2023wea}, of the conjecture presented in \cite{Navarro-Salas:2020oew} for $d=4$, which states that the fundamental invariants $\mathcal{F},\mathcal{G}$ can be fully resummed. Indeed, reducing our findings to the four-dimensional setup, one just need to recall the classical result  that contracted powers of $F_{\mu\nu}$ can be written in terms of $\mathcal{F}$, $\mathcal{G}$ alone.

As a last comment, although we have chosen the Fock--Schwinger gauge for the proof, the result is gauge independent. In fact, in the coincidence limit, the heat kernel coefficients  are made of geometric quantities; in the present case they depend only on the field strength tensor $F_{\mu\nu}$ and its covariant derivatives, which are manifestly gauge invariant.

\section{Generalisations}

\subsection{A toy model}
The operators associated with our system of a spinor in an electromagnetic background take the schematic form
\begin{equation}
    \mathcal{Q}_{\rm sch} = -\partial^2+N^\mu\partial_\mu+\alpha+\beta_\mu\bar{x}^\mu+\frac{1}{4}\gamma^2_{\mu\nu}\bar{x}^\mu\bar{x}^\nu\,.
\end{equation}
So far, throughout the discussion we have taken advantage of the fact that the associated heat kernel is closely related to Eq.~\eqref{KFc}.  Ultimately, this approach works because, in these cases, the coefficient $N^\mu$ can be written as $N^\mu = \bar{x}_\nu Y^{\nu\mu}$, where $Y$ is an antisymmetric tensor, and $\beta_\mu$ vanishes. As a consequence, the heat kernel obtained for vanishing $N^\mu$ is protected by the antisymmetry of $Y^{\mu\nu}$ against the generation of new covariant contributions.

However, when extending these resummation techniques to more general systems, it is essential to study the case where $N^\mu$ cannot be handled in such a way. To that effect, consider the following toy model for which
\begin{equation}\label{QN}
    \mathcal{Q}_{\rm N} = -\partial^2+N^\mu\partial_\mu+\alpha\,,
\end{equation}
with $N^\mu$ an arbitrary vector. Such an operator serves as a first approximation to a variety of theories, including the interaction of spinors with an axial potential~\cite{copinger} or torsion~\cite{Shapiro:2001rz}, as well as non-abelian gauge field models. The heat kernel resummation formula we propose for this case takes the form
\begin{equation}\label{HKN}
    K_{\rm N}(x,x';\tau) := \frac{1}{(4\pi\tau)^{\dime/2}}e^{-\tau\alpha-\frac{1}{4\tau}(\tau N(x')-\bar{x})^2}\Omega_{\rm N}(x,x';\tau)\,.
\end{equation}

We may justify this formula by first considering the case where $N^\mu$ is a constant vector $N_0^\mu$ (which is for example physically motivated by the Schwinger effect in the presence of constant Lorentz-violating background fields \cite{Costa:2021zkq}). This allows us to solve the heat kernel equation by finding its associated propagator, which satisfies
\begin{equation}
    \mathcal{Q}_{\rm N_0}G(x,x') = \left(-\partial^2+N_0^\mu\partial_\mu+\alpha\right)G(x,x') = \delta(x,x')\,.
\end{equation}
By performing a Fourier transform to momentum space, we can explicitly solve
\begin{widetext}
\begin{align}\label{prop}
    (p^2+iN_0^\mu p_\mu+\alpha)G(p)  = 1   \Longrightarrow G(p) = (p^2+iN_0^\mu p_\mu+\alpha)^{-1} = \int_0^{\infty} {\rm d}\tau e^{-\tau (p^2+iN_0^\mu p_\mu+\alpha)}\,.
\end{align}
\end{widetext}
By comparison with the heat kernel equation, one can straightforwardly check that the integrand in Eq.~\eqref{prop} is the Fourier transform of the heat kernel, $K_{\rm N_0}(p;\tau)$, and derive
\begin{align}
    K_{\rm N_0}(x,x';\tau) =& \int\frac{{\rm d}^\dime p}{(2\pi)^\dime} e^{ip_\mu\bar{x}^\mu}K_{\rm N_0}(p;\tau) =\nonumber \\
  & = \frac{1}{(4\pi\tau)^{d/2}}e^{-\tau\alpha-\frac{1}{4\tau}(\tau N_0-\bar{x})^2}.
\end{align}

When $N^\mu$ is not constant, we can nonetheless perform an expansion around $x'$
\begin{equation}
    N^\mu(x) = \sum_{k=0}^{\infty} \bar{x}^{\nu_1}...\bar{x}^{\nu_k} \partial_{\nu_1}...\partial_{\nu_k}N^\mu(x') = N^\mu(x') + O(\bar{x}).
\end{equation}
Introducing $K_{\rm N}$ into its defining equation and organizing all terms by their powers of the proper time $\tau$ will then yield a recurrence relation 
\begin{align}
    \begin{split}
    &\left[j+\bar{x}^\mu\left(\partial_\mu - \frac {1}{2} \big(N-N(x')\big)_\mu\right)\right]a^{(\rm N)}_j(x,x') \\
    &\hspace{0cm}=\left[\big(N(x')-N\big)_\mu\left(\partial+ \frac{1}{2}N(x')\right)^\mu + \partial^2\right]a^{(\rm N)}_{j-1}(x,x'),
    \end{split}
\end{align}
Following the same reasoning as in the previous section, one can show that the coefficients of the proper-time expansion do not depend on any invariants of the form
\begin{equation}
    \mathcal{K}_{5} = \{(N^\mu N_\mu)^j, \quad j>0\}.
\end{equation}
Explicit calculations of the first few coefficients have been presented in \cite{Franchino-Vinas:2024jvc}.

\subsection{Coupling to torsion}\label{sec:torsion}
It is worth exploring whether these results can be further generalised to more complex systems. In doing so, however, we find that such generalisations cannot be performed naively and require an attentive consideration. To exemplify this, let us consider a system consisting of a quantum spinor field interacting with an axial vector field $S^\mu$, which does not need to be a gauge field. This kind of system has been of interest for the study of some axion models \cite{Domcke:2749416}, while also serving as a gateway to understanding the interaction between spinors and torsion in more general setups~\cite{Shapiro:2001rz}. The action for such a system is given by
\begin{align}
    S_{\rm tor}:&= -\int {\rm d}^\dime x~\bar{\Psi}\left(\gamma^{\mu} D^{\rm tor}_\mu + m\right)\Psi\,, 
    \\
    D^{\rm tor}_\mu :&= \partial_{\mu} + i\eta\gamma_5S_\mu\,,
\end{align}
where $\eta$ is a (pseudoscalar) coupling constant and $\gamma_5$ is the chiral element associated with the $\gamma$ matrices (which in $\dime = 4$ is usually defined as $\gamma_5 := -i\gamma^0\gamma^1\gamma^2\gamma^3$). The associated operator we wish to study is then
\begin{align}\label{Q-S}
    \mathcal{Q}_{\rm tor} :& = \left(\gamma^\mu D^{\rm tor}_\mu + m\right)\left(-\gamma^\nu D^{\rm tor}_\nu + m\right)\nonumber \\ 
    &  = -\partial^2 + 2i\eta\gamma_5\sigma^{\mu\nu}S_\mu\partial_\nu + m^2 - \eta^2S^2 \nonumber\\
    &\hspace{2.5cm}- i\eta\gamma_5\partial_\mu S^\mu - \frac{i}{2}\eta\sigma^{\mu\nu}\mathcal{S}_{\mu\nu}\,,
\end{align}
where we define $\mathcal{S}_{\mu\nu} := \partial_\mu S_\nu - \partial_\nu S_\mu$ in analogy to the definition of $F_{\mu\nu}$. Limiting ourselves to the case where $S^\mu$ is constant, we see that the operator reduces to an expression resembling (\ref{QN}), with
\begin{align}
    \alpha & :=m^2-\eta^2 S^2\\
    N^\mu & := 2i\eta\gamma_5\sigma^{\nu\mu}S_\nu\,.
\end{align}
Unlike in the models of the previous section, this time $N^\mu$ is not a mere constant \textit{scalar-valued} vector but a constant \textit{matrix-valued} vector, reflecting its nontrivial spinor structure from the $\gamma$-matrices product.
The specific structure of $N^\mu$ can be used to show that
\begin{equation}
    \{N^\mu, N^\nu\} = 8\eta^2\left(S^\mu S^\nu - S^2 \eta^{\mu\nu}\right)\mathbb{I},
\end{equation}
allowing us to search for an explicit solution to the heat kernel equation in a manner similar to the one used before. The result obtained,
\begin{equation}
    K_{\rm tor}(x,x';\tau) = i e^{-\tau\alpha} \int\dfrac{{\rm d}^d p}{(2\pi)^\dime} e^{-\tau p^2 + ip\cdot\bar{x}} \exp(\tau N^\mu p_\mu),
\end{equation}
presents a major difference with respect to the case of a scalar-valued vector $N^\mu$: the factor $\exp(\tau N^\mu p_\mu)$ is no longer the exponential of a scalar (which would allow the integral to be evaluated straightforwardly), but rather the exponential of a nontrivial matrix object. The integral can be explicitely done by performing a formal power series expansion and then reorganising the outcome. So far it has proven nontrivial to find a closed form for this heat kernel. In its current form, it is still possible to show that the resulting heat kernel proper-time expansion will once again effectively resum all the contributions from $S_\mu S^\mu$ (appearing as an effective curvature / mass correction term), but it remains to be seen whether further results can be derived.

\section{Conclusions}

The development of methods capable of probing quantum effects beyond perturbation theory remains a central pursuit in theoretical and mathematical physics. A wide range of strategies have been employed, ranging from lattice formulations~\cite{Lattice:2023} to semiclassical and instanton techniques~\cite{DunneHall:1998,DunneWang:2006,Schutzhold:2008, DegliEsposti:2024upq, Semren:2025dix}, which capture nonperturbative contributions via tunnelling configurations and saddle points of the Euclidean action. Alongside these, effective-action and heat kernel methods~\cite{Parker_Toms_2009} provide a complementary framework, in which nonperturbative information can be extracted from the spectral properties of differential operators. In particular, while the standard short-time expansion of the heat kernel yields an asymptotic perturbative series, appropriate resummation techniques can reveal analytic structures that encode  nonperturbative physics. Unlike the case of instanton methods, where analytic results can be obtained just for certain models, our results are rather general, depending only on the assumption of large fields.

Within the broader framework of quantum field theory in curved spacetime, such techniques acquire special significance. Originally conceived as an intermediate step toward a quantised theory of gravity, this field has evolved into a rich discipline in its own right, revealing phenomena that illuminate the interplay between gravitation and quantum mechanics across energy scales. The Hawking effect~\cite{Birrell_Davies_1982}, which predicts spontaneous particle creation in strong gravitational fields, stands as its most emblematic example. Even in the simplest settings (for example, when a single quantum field interacts with a classical background) substantial conceptual and computational challenges persist. Standard approaches such as Feynman’s diagrammatic expansion~\cite{Srednicki_2007} rely on infinite perturbative series, for which the partial sums' convergence and interpretation become increasingly opaque beyond leading order, particularly in gravitational or strongly coupled regimes. 

In this context, the heat kernel formalism provides a unifying framework where semiclassical perturbative and nonperturbative effects can be explored. This perspective motivates the program developed in recent years to identify and formalise resummation patterns that capture nonperturbative information directly from the heat kernel expansion. The results presented in this paper contribute to this effort and can be summarised in three main outcomes.

First, the resummation conjecture, originally posed in $d=4$ by Navarro-Salas and Pla~\cite{Navarro-Salas:2020oew}, has now been explicitly proved for both scalar and spinor quantum electrodynamics. The  proof in the present manuscript yields a new resummed form of the heat kernel expansion for fermionic systems, effectively capturing all invariants built from fully contracted powers of the electromagnetic tensor $F_{\mu\nu}$, together with a ``mass correction'' term proportional to $\sigma^{\mu\nu}F_{\mu\nu}$. Our results generalise the original conjecture, inasmuch as they are valid in flat spaces of \textit{arbitrary} dimensions $d$, with a restriction to even-dimensional spaces for fermionic systems. The peculiarity of $d=4$ resides in the fact that any contracted power of the field strength can be written in terms of the invariants $\mathcal{F}$ and $\mathcal{G}$, which form actually the language employed in Ref.~\cite{Navarro-Salas:2020oew}.

Second, although a detailed order-by-order derivation of the proper-time expansion coefficients was not carried out here, they can be obtained recursively from Eq.~(\ref{HKrec-eq}), yielding results consistent with earlier analyses~\cite{Vassilevich:2003xt,vandeVen:1997pf,Franchino-Vinas:2024wof}. This resummed heat kernel expansion thus provides an alternative and compact tool for analysing one-loop effective actions and their associated phenomena, including particle-pair creation in the Schwinger process.

Third, the formalism developed here may admit extensions to more general settings. While such generalisations lie beyond the scope of the present work, possible directions include non-Abelian gauge theories and curved spacetime backgrounds. These cases are expected to present additional conceptual and computational challenges, whose resolution may benefit from comparisons with other frameworks designed to study loop effects, such as 
numerical approaches~\cite{Edwards:2025cco}, large-N expansions~\cite{Karbstein:2023yee}, large quantum non-linear parameter resummations~\cite{Mironov:2020gbi} and the worldline formalism~\cite{Fecit:2025kqb}. In particular, recent developments concerning axial couplings~\cite{Bastianelli:2024vkp} and gravitational setups~\cite{Ilderton:2025umd} may provide further insight into the structure of these generalisations. 

Beyond their formal aspects, resummed heat kernel methods open promising avenues for phenomenological applications. As previously discussed, the resummed kernel derived in Sec.~\ref{sec:3} reproduces the known results for the Schwinger effect and at the same time, for the scalar case, suggests the existence of analogous ``Schwinger-like'' mechanisms for particle creation. Testing these ideas could be relevant in certain inflationary models or in scenarios involving ultralight dark matter~\cite{Patt:2006fw,Piazza2010}. Such directions underline the potential of resummed heat kernel formulations not only as analytic tools for nonperturbative physics, but also as bridges connecting semiclassical field theory with observational frontiers.

\section*{Acknowledgments}
SAF thanks the members of the Institut Denis Poisson, especially M. Chernodub, for their warm hospitality. The authors acknowledge fruitful discussions and funding from the workshops ``New Trends in First Quantisation: Field Theory, Gravity and Quantum Computing'' (Heraeus Stiftung) and ``First Quantisation for Physics in Strong Fields'' (University of Edinburgh). The work of VV has been partially funded by Next Generation EU through the project ``Geometrical and Topological effects on Quantum Matter (GeTOnQuaM)''. The research activities of SAF, CGP and VV have been carried out in the framework of the INFN Research Project QGSKY. The work of SP was funded by the Deutsche Forschungsgemeinschaft (DFG, German Research Foundation) under Germany’s Excellence Strategy – EXC 2094 – 390783311.
SAF and FDM acknowledge the support from Consejo Nacional de Investigaciones Científicas y Técnicas (CONICET) through Project PIP 11220200101426CO.
SAF acknowledges the support of UNLP through Project 11/X748. 
The authors would like to acknowledge the contribution of the COST Action CA23130.
The authors also extend their appreciation to the Italian National Group of Mathematical Physics (GNFM, INdAM) for its support.

\bibliography{bibliografia} 

@article{Navarro-Salas:2020oew,
    author = "Navarro-Salas, Jose and Pla, Silvia",
    title = "{$(\mathcal{F},\mathcal{G})$-summed form of the QED effective action}",
    eprint = "2011.09743",
    archivePrefix = "arXiv",
    primaryClass = "hep-th",
    doi = "10.1103/PhysRevD.103.L081702",
    journal = "Phys. Rev. D",
    volume = "103",
    number = "8",
    pages = "L081702",
    year = "2021"
}

@article{Franchino-Vinas:2023wea,
    author = "Franchino-Vi{\~n}as, S. A. and Garc{\'\i}a-P{\'e}rez, C. and Mazzitelli, F. D. and Vitagliano, V. and Haimovichi, U. Wainstein",
    title = "{Resummed heat kernel and effective action for Yukawa and QED}",
    eprint = "2312.16303",
    archivePrefix = "arXiv",
    primaryClass = "hep-th",
    doi = "10.1016/j.physletb.2024.138684",
    journal = "Phys. Lett. B",
    volume = "854",
    pages = "138684",
    year = "2024"
}

@inproceedings{Franchino-Vinas:2024jvc,
    author = "Franchino-Vi{\~n}as, S. A. and Garc{\'\i}a-P{\'e}rez, C. and Mazzitelli, F. D. and Pla, S. and Vitagliano, V. and Wainstein-Haimovichi, U.",
    title = "{Strong-field resummed heat kernels and effective actions: inhomogeneous fields}",
    booktitle = "{17th Marcel Grossmann Meeting}: {On Recent Developments in Theoretical and Experimental General Relativity, Gravitation, and Relativistic Field Theories}",
    eprint = "2410.11364",
    archivePrefix = "arXiv",
    primaryClass = "hep-th",
    month = "10",
    year = "2024"
}

@book{pascual1984qcd,
  title={QCD: Renormalization for the Practitioner},
  author={Pascual, Pedro and Tarrach, Rolf},
  year={1984},
  publisher={Springer}
}

@article{Shapiro:2001rz,
    author = "Shapiro, I. L.",
    title = "{Physical aspects of the space-time torsion}",
    eprint = "hep-th/0103093",
    archivePrefix = "arXiv",
    reportNumber = "DF-UFJF-01-04",
    doi = "10.1016/S0370-1573(01)00030-8",
    journal = "Phys. Rept.",
    volume = "357",
    pages = "113",
    year = "2002"
}

@article{vandeVen:1997pf,
    author = "van de Ven, Anton E. M.",
    title = "{Index free heat kernel coefficients}",
    eprint = "hep-th/9708152",
    archivePrefix = "arXiv",
    reportNumber = "DESY-97-162, DESY-97-126",
    doi = "10.1088/0264-9381/15/8/014",
    journal = "Class. Quant. Grav.",
    volume = "15",
    pages = "2311--2344",
    year = "1998"
}

@article{copinger,
  title = {Euler-Heisenberg Lagrangian under an axial gauge field},
  author = {Copinger, Patrick and Hattori, Koichi and Yang, Di-Lun},
  journal = {Phys. Rev. D},
  volume = {107},
  issue = {5},
  pages = {056016},
  numpages = {18},
  year = {2023},
  month = {Mar},
  publisher = {American Physical Society},
  doi = {10.1103/PhysRevD.107.056016},
  url = {https://link.aps.org/doi/10.1103/PhysRevD.107.056016}
}

@article{Domcke:2749416,
      author        = "Domcke, Valerie and Ema, Yohei and Mukaida, Kyohei",
      title         = "{Axion assisted Schwinger effect}",
      archivePrefix = "arXiv",
      eprint        = "2101.05192",
      reportNumber  = "DESY-21-006, CERN-TH-2021-010",
      journal       = "JHEP",
      volume        = "2105",
      pages         = "001",
      year          = "2021",
      url           = "https://cds.cern.ch/record/2749416",
      doi           = "10.1007/JHEP05(2021)001",
}

@article{DunneHall:1998,
  title = {QED effective action in time dependent electric backgrounds},
  author = {Dunne, Gerald and Hall, Theodore},
  journal = {Phys. Rev. D},
  volume = {58},
  issue = {10},
  pages = {105022},
  numpages = {13},
  year = {1998},
  month = {Oct},
  publisher = {American Physical Society},
  doi = {10.1103/PhysRevD.58.105022},
  url = {https://link.aps.org/doi/10.1103/PhysRevD.58.105022}
}

@article{DunneWang:2006,
  title = {Worldline instantons and the fluctuation prefactor},
  author = {Dunne, Gerald V. and Wang, Qing-hai and Gies, Holger and Schubert, Christian},
  journal = {Phys. Rev. D},
  volume = {73},
  issue = {6},
  pages = {065028},
  numpages = {13},
  year = {2006},
  month = {Mar},
  publisher = {American Physical Society},
  doi = {10.1103/PhysRevD.73.065028},
  url = {https://link.aps.org/doi/10.1103/PhysRevD.73.065028}
}

@article{Schutzhold:2008,
  title = {Dynamically Assisted Schwinger Mechanism},
  author = {Sch\"utzhold, Ralf and Gies, Holger and Dunne, Gerald},
  journal = {Phys. Rev. Lett.},
  volume = {101},
  issue = {13},
  pages = {130404},
  numpages = {4},
  year = {2008},
  month = {Sep},
  publisher = {American Physical Society},
  doi = {10.1103/PhysRevLett.101.130404},
  url = {https://link.aps.org/doi/10.1103/PhysRevLett.101.130404}
}

@article{Lattice:2023,
  title = {Lattice calculation of the short and intermediate time-distance hadronic vacuum polarization contributions to the muon magnetic moment using twisted-mass fermions},
  author = {Alexandrou, C. and Bacchio, S. and Dimopoulos, P. and Finkenrath, J. and Frezzotti, R. and Gagliardi, G. and Garofalo, M. and Hadjiyiannakou, K. and Kostrzewa, B. and Jansen, K. and Lubicz, V. and Petschlies, M. and Sanfilippo, F. and Simula, S. and Urbach, C. and Wenger, U.},
  collaboration = {Extended Twisted Mass Collaboration},
  journal = {Phys. Rev. D},
  volume = {107},
  issue = {7},
  pages = {074506},
  numpages = {42},
  year = {2023},
  month = {Apr},
  publisher = {American Physical Society},
  doi = {10.1103/PhysRevD.107.074506},
  url = {https://link.aps.org/doi/10.1103/PhysRevD.107.074506}
}

@article{Piazza2010,
  title = {Sub-eV scalar dark matter through the super-renormalizable Higgs portal},
  author = {Piazza, Federico and Pospelov, Maxim},
  journal = {Phys. Rev. D},
  volume = {82},
  issue = {4},
  pages = {043533},
  numpages = {7},
  year = {2010},
  month = {Aug},
  publisher = {American Physical Society},
  doi = {10.1103/PhysRevD.82.043533},
  url = {https://link.aps.org/doi/10.1103/PhysRevD.82.043533}
}

@book{Birrell_Davies_1982,
place={Cambridge},
series={Cambridge Monographs on Mathematical Physics},
title={Quantum Fields in Curved Space},
publisher={Cambridge University Press},
author={Birrell, N. D. and Davies, P. C. W.},
year={1982},
collection={Cambridge Monographs on Mathematical Physics}}

@book{Parker_Toms_2009, place={Cambridge}, series={Cambridge Monographs on Mathematical Physics}, title={Quantum Field Theory in Curved Spacetime: Quantized Fields and Gravity}, publisher={Cambridge University Press}, author={Parker, Leonard and Toms, David}, year={2009}, collection={Cambridge Monographs on Mathematical Physics}}

@book{Srednicki_2007, place={Cambridge}, title={Quantum Field Theory}, publisher={Cambridge University Press}, author={Srednicki, Mark}, year={2007}}

@article{Fecit:2025kqb,
    author = "Fecit, Filippo and Franchino-Vi{\~n}as, Sebasti{\'a}n A. and Mazzitelli, Francisco D.",
    title = "{Resummed effective actions and heat kernels: the Worldline approach and Yukawa assisted pair creation}",
    eprint = "2501.17094",
    archivePrefix = "arXiv",
    primaryClass = "hep-th",
    doi = "10.1007/JHEP07(2025)041",
    journal = "JHEP",
    volume = "07",
    pages = "041",
    year = "2025"
}

@article{Bastianelli:2024vkp,
    author = "Bastianelli, F. and Corradini, O. and Edwards, J. P. and McKeon, D. G. C. and Schubert, C.",
    title = "{Unified worldline treatment of Yukawa and axial couplings}",
    eprint = "2406.19988",
    archivePrefix = "arXiv",
    primaryClass = "hep-th",
    doi = "10.1007/JHEP11(2024)152",
    journal = "JHEP",
    volume = "11",
    pages = "152",
    year = "2024"
}

@book{Coleman1985,
  author = {S. Coleman},
  title = {Aspects of Symmetry},
  year = {1985},
  publisher = {Cambridge University Press}
}

@article{Callan:1977pt,
    author = "Callan, Jr., Curtis G. and Coleman, Sidney R.",
    title = "{The Fate of the False Vacuum. 2. First Quantum Corrections}",
    reportNumber = "HUTP-77-A032",
    doi = "10.1103/PhysRevD.16.1762",
    journal = "Phys. Rev. D",
    volume = "16",
    pages = "1762--1768",
    year = "1977"
}

@article{Belavin:1975fg,
    author = "Belavin, A. A. and Polyakov, Alexander M. and Schwartz, A. S. and Tyupkin, Yu. S.",
    editor = "Taylor, J. C.",
    title = "{Pseudoparticle Solutions of the Yang-Mills Equations}",
    doi = "10.1016/0370-2693(75)90163-X",
    journal = "Phys. Lett. B",
    volume = "59",
    pages = "85--87",
    year = "1975"
}

@book{Gilkey1995,
  author = {P. B. Gilkey},
  title = {Invariance Theory, the Heat Equation and the Atiyah-Singer Index Theorem},
  year = {1995},
  publisher = {CRC Press}
}

@article{Dunne:2012ae,
    author = "Dunne, Gerald V. and Unsal, Mithat",
    title = "{Resurgence and Trans-series in Quantum Field Theory: The CP(N-1) Model}",
    eprint = "1210.2423",
    archivePrefix = "arXiv",
    primaryClass = "hep-th",
    doi = "10.1007/JHEP11(2012)170",
    journal = "JHEP",
    volume = "11",
    pages = "170",
    year = "2012"
}

@article{Voros:1986vw,
    author = "Voros, A.",
    title = "{Spectral Functions, Special Functions and Selberg Zeta Function}",
    reportNumber = "SACLAY-PhT-86-114",
    doi = "10.1007/BF01212422",
    journal = "Commun. Math. Phys.",
    volume = "110",
    pages = "439",
    year = "1987"
}

@article{Avramidi:2009quh,
    author = "Avramidi, Ivan G. and Fucci, Guglielmo",
    title = "{Non-perturbative Heat Kernel Asymptotics on Homogeneous Abelian Bundles}",
    eprint = "0810.4889",
    archivePrefix = "arXiv",
    primaryClass = "math-ph",
    doi = "10.1007/s00220-009-0804-6",
    journal = "Commun. Math. Phys.",
    volume = "291",
    pages = "543--577",
    year = "2009"
}

@article{Affleck:1981bma,
    author = "Affleck, Ian K. and Alvarez, Orlando and Manton, Nicholas S.",
    title = "{Pair Production at Strong Coupling in Weak External Fields}",
    reportNumber = "Print-81-0812 (PRINCETON)",
    doi = "10.1016/0550-3213(82)90455-2",
    journal = "Nucl. Phys. B",
    volume = "197",
    pages = "509--519",
    year = "1982"
}

@article{Dunne:2005sx,
    author = "Dunne, Gerald V. and Schubert, Christian",
    title = "{Worldline instantons and pair production in inhomogeneous fields}",
    eprint = "hep-th/0507174",
    archivePrefix = "arXiv",
    doi = "10.1103/PhysRevD.72.105004",
    journal = "Phys. Rev. D",
    volume = "72",
    pages = "105004",
    year = "2005"
}

@article{Parikh:1999mf,
    author = "Parikh, Maulik K. and Wilczek, Frank",
    title = "{Hawking radiation as tunneling}",
    eprint = "hep-th/9907001",
    archivePrefix = "arXiv",
    reportNumber = "PUPT-1775, SPIN-1998-12, IASSNS-HEP-98-22",
    doi = "10.1103/PhysRevLett.85.5042",
    journal = "Phys. Rev. Lett.",
    volume = "85",
    pages = "5042--5045",
    year = "2000"
}

@article{Vassilevich:2003xt,
    author = "Vassilevich, D.V.",
    title = "{Heat kernel expansion: User's manual}",
    eprint = "hep-th/0306138",
    archivePrefix = "arXiv",
    doi = "10.1016/j.physrep.2003.09.002",
    journal = "Phys. Rept.",
    volume = "388",
    pages = "279--360",
    year = "2003"
}

@article{Barvinsky:1987uw,
    author = "Barvinsky, A.O. and Vilkovisky, G.A.",
    title = "{Beyond the Schwinger-Dewitt Technique: Converting Loops Into Trees and In-In Currents}",
    doi = "10.1016/0550-3213(87)90681-X",
    journal = "Nucl. Phys. B",
    volume = "282",
    pages = "163--188",
    year = "1987"
}

@article{Costa:2021zkq,
    author = "Costa, Rafael L. J. and Sobreiro, Rodrigo F.",
    title = "{One-loop Schwinger effect in the presence of Lorentz-violating background fields}",
    eprint = "2112.11967",
    archivePrefix = "arXiv",
    primaryClass = "hep-th",
    doi = "10.1140/epjc/s10052-022-10625-1",
    journal = "Eur. Phys. J. C",
    volume = "82",
    number = "8",
    pages = "677",
    year = "2022"
}

@article{Barvinskii:2024iqz,
    author = "Barvinskii, Andrei O. and Wachowski, W.",
    title = "{Schwinger{\textemdash}DeWitt technique in quantum gravity}",
    doi = "10.3367/UFNe.2024.02.039646",
    journal = "Phys. Usp.",
    volume = "67",
    number = "8",
    pages = "751--767",
    year = "2024"
}

@book{DeWitt:1965,
    author = "DeWitt, Bryce S.",
    title = "{Dynamical Theory of Groups and Fields}",
    year = "1965",
    isbn = "0888-6105",
    publisher = "Gordon and Breach, Science Publishers Ltd."
}

@article{Flachi:2025gxo,
    author = "Flachi, Antonino and Tanaka, Takahiro",
    title = "{Quantum droplets in curved space}",
    eprint = "2507.13682",
    archivePrefix = "arXiv",
    primaryClass = "hep-th",
    month = "7",
    year = "2025"
}

@article{Edwards:2021cyp,
    author = "Edwards, James P. and Gonz\'alez-Dom\'inguez, V\'ictor A. and Huet, Idrish and Trejo, Mar\'ia Anabel",
    title = "{Resummation for quantum propagators in bounded spaces}",
    eprint = "2110.04969",
    archivePrefix = "arXiv",
    primaryClass = "quant-ph",
    doi = "10.1103/PhysRevE.105.064132",
    journal = "Phys. Rev. E",
    volume = "105",
    number = "6",
    pages = "064132",
    year = "2022"
}

@article{Flachi:2015sva,
    author = "Flachi, Antonino and Fukushima, Kenji and Vitagliano, Vincenzo",
    title = "{Geometrically induced magnetic catalysis and critical dimensions}",
    eprint = "1502.06090",
    archivePrefix = "arXiv",
    primaryClass = "hep-th",
    doi = "10.1103/PhysRevLett.114.181601",
    journal = "Phys. Rev. Lett.",
    volume = "114",
    number = "18",
    pages = "181601",
    year = "2015"
}

@article{Castro:2018iqt,
    author = "Castro, Eduardo V. and Flachi, Antonino and Ribeiro, Pedro and Vitagliano, Vincenzo",
    title = "{Symmetry Breaking and Lattice Kirigami}",
    eprint = "1803.09495",
    archivePrefix = "arXiv",
    primaryClass = "hep-th",
    doi = "10.1103/PhysRevLett.121.221601",
    journal = "Phys. Rev. Lett.",
    volume = "121",
    number = "22",
    pages = "221601",
    year = "2018"
}

@article{Flachi:2019btk,
    author = "Flachi, Antonino and Vitagliano, Vincenzo",
    title = "{Symmetry breaking and lattice kirigami: finite temperature effects}",
    eprint = "1904.06912",
    archivePrefix = "arXiv",
    primaryClass = "hep-th",
    doi = "10.1103/PhysRevD.99.125010",
    journal = "Phys. Rev. D",
    volume = "99",
    number = "12",
    pages = "125010",
    year = "2019"
}

@article{Flachi:2014jra,
    author = "Flachi, Antonino and Fukushima, Kenji",
    title = "{Chiral Mass-Gap in Curved Space}",
    eprint = "1406.6548",
    archivePrefix = "arXiv",
    primaryClass = "hep-th",
    doi = "10.1103/PhysRevLett.113.091102",
    journal = "Phys. Rev. Lett.",
    volume = "113",
    number = "9",
    pages = "091102",
    year = "2014"
}

@article{Brown:1975bc,
    author = "Brown, M. R. and Duff, M. J.",
    title = "{Exact Results for Effective Lagrangians}",
    doi = "10.1103/PhysRevD.11.2124",
    journal = "Phys. Rev. D",
    volume = "11",
    pages = "2124--2135",
    year = "1975"
}

@article{Heisenberg:1936nmg,
    author = "Heisenberg, W. and Euler, H.",
    title = "{Consequences of Dirac's theory of positrons}",
    eprint = "physics/0605038",
    archivePrefix = "arXiv",
    doi = "10.1007/BF01343663",
    journal = "Z. Phys.",
    volume = "98",
    number = "11-12",
    pages = "714--732",
    year = "1936"
}

@article{Weisskopf:1936hya,
    author = "Weisskopf, V.",
    title = "{The electrodynamics of the vacuum based on the quantum theory of the electron}",
    journal = "Kong. Dan. Vid. Sel. Mat. Fys. Med.",
    volume = "14N6",
    number = "6",
    pages = "1--39",
    year = "1936"
}

@article{Barvinsky:1990up,
    author = "Barvinsky, A. O. and Vilkovisky, G. A.",
    title = "{Covariant perturbation theory. 2: Second order in the curvature. General algorithms}",
    doi = "10.1016/0550-3213(90)90047-H",
    journal = "Nucl. Phys. B",
    volume = "333",
    pages = "471--511",
    year = "1990"
}

@article{Calzetta:1986pj,
    author = "Calzetta, E. and Jack, I. and Parker, L.",
    title = "{Quantum gauge fields at high curvature}",
    doi = "10.1103/PhysRevD.33.953",
    journal = "Phys. Rev. D",
    volume = "33",
    pages = "953--977",
    year = "1986"
}

@article{Jack:1985mw,
    author = "Jack, Ian and Parker, Leonard",
    title = "{Proof of Summed Form of Proper Time Expansion for Propagator in Curved Space-time}",
    reportNumber = "Print-85-0139 (WISCONSIN)",
    doi = "10.1103/PhysRevD.31.2439",
    journal = "Phys. Rev. D",
    volume = "31",
    pages = "2439",
    year = "1985"
}

@article{Parker:1984dj,
    author = "Parker, Leonard and Toms, David J.",
    title = "{New Form for the Coincidence Limit of the Feynman Propagator, or Heat Kernel, in Curved Space-time}",
    reportNumber = "Print-84-0994 (WISCONSIN)",
    doi = "10.1103/PhysRevD.31.953",
    journal = "Phys. Rev. D",
    volume = "31",
    pages = "953",
    year = "1985"
}

@article{Ahmadiniaz:2022bwy,
    author = "Ahmadiniaz, N. and Franchino-Vi\~nas, S. A. and Manzo, L. and Mazzitelli, F. D.",
    title = "{Local Neumann semitransparent layers: Resummation, pair production, and duality}",
    eprint = "2208.07383",
    archivePrefix = "arXiv",
    primaryClass = "hep-th",
    doi = "10.1103/PhysRevD.106.105022",
    journal = "Phys. Rev. D",
    volume = "106",
    number = "10",
    pages = "105022",
    year = "2022"
}

@article{Franchino-Vinas:2020okl,
    author = "Franchino-Vi\~nas, S. A. and Mazzitelli, F. D.",
    title = "{Effective action for delta potentials: spacetime-dependent inhomogeneities and Casimir self-energy}",
    eprint = "2010.11144",
    archivePrefix = "arXiv",
    primaryClass = "hep-th",
    doi = "10.1103/PhysRevD.103.065006",
    journal = "Phys. Rev. D",
    volume = "103",
    number = "6",
    pages = "065006",
    year = "2021"
}

@article{Franchino-Vinas:2018gzr,
    author = "Franchino-Vi\~nas, Sebasti\'an A. and de Paula Netto, Tib\'erio and Shapiro, Ilya L. and Zanusso, Omar",
    title = "{Form factors and decoupling of matter fields in four-dimensional gravity}",
    eprint = "1812.00460",
    archivePrefix = "arXiv",
    primaryClass = "hep-th",
    doi = "10.1016/j.physletb.2019.01.021",
    journal = "Phys. Lett. B",
    volume = "790",
    pages = "229--236",
    year = "2019"
}

@article{Guven:1986gi,
    author = "Guven, Jemal",
    title = "{An Expansion for the Effective Action of an Interacting Quantum Field Theory in Curved Space}",
    reportNumber = "MIT-CTP-1411",
    doi = "10.1103/PhysRevD.35.2378",
    journal = "Phys. Rev. D",
    volume = "35",
    pages = "2378",
    year = "1987"
}

@article{Avramidi:1995ik,
    author = "Avramidi, I. G.",
    title = "{Covariant algebraic method for calculation of the low-energy heat kernel}",
    eprint = "hep-th/9503132",
    archivePrefix = "arXiv",
    doi = "10.1063/1.531371",
    journal = "J. Math. Phys.",
    volume = "36",
    pages = "5055--5070",
    year = "1995"
}

@article{Wigner:1932eb,
    author = "Wigner, Eugene P.",
    title = "{On the quantum correction for thermodynamic equilibrium}",
    doi = "10.1103/PhysRev.40.749",
    journal = "Phys. Rev.",
    volume = "40",
    pages = "749--760",
    year = "1932"
}

@article{Gorbar:2003yt,
    author = "Gorbar, Eduard V. and Shapiro, Ilya L.",
    title = "{Renormalization group and decoupling in curved space. 2. The Standard model and beyond}",
    eprint = "hep-ph/0303124",
    archivePrefix = "arXiv",
    reportNumber = "DF-UFJF-03-01",
    doi = "10.1088/1126-6708/2003/06/004",
    journal = "JHEP",
    volume = "06",
    pages = "004",
    year = "2003"
}

@article{Gorbar:2002pw,
    author = "Gorbar, E. V. and Shapiro, I. L.",
    title = "{Renormalization group and decoupling in curved space}",
    eprint = "hep-ph/0210388",
    archivePrefix = "arXiv",
    reportNumber = "DF-UFJF-02-02",
    doi = "10.1088/1126-6708/2003/02/021",
    journal = "JHEP",
    volume = "02",
    pages = "021",
    year = "2003"
}

@article{Silva:2023lts,
    author = "Silva, Wagno Cesar e and Shapiro, Ilya L.",
    title = "{Effective approach to the Antoniadis-Mottola model: quantum decoupling of the higher derivative terms}",
    eprint = "2301.13291",
    archivePrefix = "arXiv",
    primaryClass = "hep-th",
    doi = "10.1007/JHEP07(2023)097",
    journal = "JHEP",
    volume = "07",
    pages = "097",
    year = "2023"
}

@article{Patt:2006fw,
    author = "Patt, Brian and Wilczek, Frank",
    title = "{Higgs-field portal into hidden sectors}",
    eprint = "hep-ph/0605188",
    archivePrefix = "arXiv",
    reportNumber = "MIT-CTP-3745",
    month = "5",
    year = "2006"
}

@article{Franchino-Vinas:2024wof,
    author = "Franchino-Vi\~nas, S. A.",
    title = "{Comment on ``Index-free heat kernel coefficients''}",
    eprint = "2401.01296",
    archivePrefix = "arXiv",
    primaryClass = "hep-th",
    doi = "10.1088/1361-6382/ad4949",
    journal = "Class. Quant. Grav.",
    volume = "41",
    number = "12",
    pages = "128001",
    year = "2024"
}

@article{Semren:2025dix,
    author = "Semr{\'e}n, Philip and Torgrimsson, Greger",
    title = "{Worldline instantons for nonperturbative particle production by space and time dependent gravitational fields}",
    eprint = "2508.01901",
    archivePrefix = "arXiv",
    primaryClass = "hep-th",
    month = "8",
    year = "2025"
}

@article{DegliEsposti:2024upq,
    author = "Degli Esposti, Gianluca and Torgrimsson, Greger",
    title = "{Schwinger pair production in spacetime fields: Moir{\'e} patterns, Aharonov-Bohm phases, and Sturm-Liouville eigenvalues}",
    eprint = "2412.19709",
    archivePrefix = "arXiv",
    primaryClass = "hep-ph",
    doi = "10.1103/3143-w4w3",
    journal = "Phys. Rev. D",
    volume = "112",
    number = "1",
    pages = "016026",
    year = "2025"
}

@article{Ilderton:2025umd,
    author = "Ilderton, Anton and Rajeev, Karthik",
    title = "{Tunnelling amplitudes and Hawking radiation from worldline QFT}",
    eprint = "2508.00997",
    archivePrefix = "arXiv",
    primaryClass = "hep-th",
    month = "8",
    year = "2025"
}

@article{Karbstein:2023yee,
    author = "Karbstein, Felix",
    title = "{Towards the full Heisenberg-Euler effective action at large N}",
    eprint = "2312.09804",
    archivePrefix = "arXiv",
    primaryClass = "hep-th",
    doi = "10.1007/JHEP02(2024)180",
    journal = "JHEP",
    volume = "02",
    pages = "180",
    year = "2024"
}

@article{Edwards:2025cco,
    author = {Edwards, James P. and Ahmadiniaz, Naser and Schmidt, Sebastian M. and Kohlf{\"u}rst, Christian},
    title = "{Relativistic quantum kinetic theory: Higher order contributions in assisted Schwinger pair production}",
    eprint = "2502.14703",
    archivePrefix = "arXiv",
    primaryClass = "hep-ph",
    doi = "10.1103/mgst-vn3c",
    journal = "Phys. Rev. D",
    volume = "112",
    number = "3",
    pages = "L031901",
    year = "2025"
}

@article{Mironov:2020gbi,
    author = "Mironov, A. A. and Meuren, S. and Fedotov, A. M.",
    title = "{Resummation of QED radiative corrections in a strong constant crossed field}",
    eprint = "2003.06909",
    archivePrefix = "arXiv",
    primaryClass = "hep-th",
    doi = "10.1103/PhysRevD.102.053005",
    journal = "Phys. Rev. D",
    volume = "102",
    number = "5",
    pages = "053005",
    year = "2020"
}

\end{document}